\documentclass[preprint,prb]{revtex4-1}

\usepackage[latin1]{inputenc}
\usepackage{amsmath}
\usepackage{amssymb}
\usepackage{epsfig}
\usepackage{mathrsfs}

\begin{document}
\title{Strong Purcell effect observed in single thick shell CdSe/CdS nanocrystals coupled to localized surface plasmons}
\author{D. Canneson$^\ddag$}
\author{I. Mallek-Zouari$^\ddag$}
\author{S. Buil$^\ddag$}
\author{X. Qu\'elin$^\ddag$}
\author{C. Javaux$^{\Diamond}$}
\author{B. Mahler$^{\Diamond}$}
\author{B. Dubertret$^{\Diamond}$}
\author{J.-P. Hermier$^\ddag,^{\odot}$}
\affiliation{$^\ddag$Groupe d'\'{E}tude de la Mati\`ere Condens\'ee, Universit\'e de Versailles-Saint-Quentin-en-Yvelines, CNRS UMR8635, 45 avenue des \'{E}tats-Unis, 78035 Versailles, France}
\affiliation{$^{\Diamond}$Laboratoire de Physique et d'\'Etude des Mat\'eriaux, CNRS UMR8213, ESPCI, 10 rue Vauquelin, 75231 Paris, France}
\affiliation{$^{\odot}$Institut Universitaire de France}

\begin{abstract}
High quality factor dielectric cavities designed to a nanoscale accuracy are mostly used to increase the spontaneous emission rate of a single emitter. Here we show that the coupling, at room temperature, between thick shell CdSe/CdS nanocrystals and random metallic films offers a very promising alternative approach. Optical modes confined at the nanoscale induce strong Purcell factors reaching values as high as 60. Moreover the quantum emission properties can be tailored: strong antibunching or radiative biexcitonic cascades can be obtained with high photon collection efficiency and extremely reduced blinking. 
\end{abstract}

        
\maketitle   
Colloidal core-shell nanocrystals (NCs) are very promising emitters for a wide range of applications such as quantum information \cite{Brokmann04}, optoelectronics \cite{Talapin10} and biological labeling \cite{Michalet05}. They are photostable at room temperature and present high quantum efficiency (QE). Recently, the synthesis of novel CdSe/CdS NCs, characterized by a very thick shell, resulted in drastically reducing their blinking \cite{Mahler08,Chen08}. 

This result also illustrates a means to modify the fluorescence of a single emitter. Besides the modification of the fluorophore itself, it can be placed in a controlled electromagnetic environment. This strategy has been widely implemented using cavities. In particular, the spontaneous emission rate can be controlled according to the Purcell approach \cite{Purcell46}, with important implications in the field of quantum optics. 

Dielectric cavities with high quality factor $Q$ have been widely investigated \cite{Gerard98,Badolato05,Hennessy07}. The Purcell factor (defined as the increase of the spontaneous decay rate) is proportional to $Q$/$V$, where $V$ is the modal volume, when the width of the emitter resonance is much smaller than that of the dielectric cavity considered without losses. The optimization of optical cavities has consisted in decreasing $V$ and increasing $Q$. However, when $Q_e$, the quality factor of the emitter ($Q_e = \lambda/\Delta \lambda$ where $\lambda$ is the emission wavelength and $\Delta\lambda$ is the linewidth) is very much lower than $Q$, it replaces $Q$ in the $F_P$ formula. Increasing $Q$ has thus no more effect. The use of plasmonic structures to obtain a high $F_P$ factor with a large bandwidth, is very promising. In this case, the key to high $F_P$, even if $F_P$ cannot be rigorously defined from $Q$ and $V$ \cite{Koenderink10}, is light confinement \cite{Akimov07,Orrit07,Esteban10}.

The most common approach to improve cavities has consisted in making well ordered structures which require nanoscale engineering \cite{Gerard98,Badolato05,Hennessy07}. An alternative strategy, based on Anderson localized modes obtained by a controlled disorder in a photonic crystal \cite{Sapienza10}, resulted in simultaneously high $Q$ value and very low $V$ value ($\sim \lambda^3$). With the use of a metallic film, the disorder leads to even stronger confinement, at the nanometer scale \cite{Ducourtieux01,Stockman01,Krachmalnicoff10}.

Here, we demonstrate the very promising opportunities presented by the coupling of thick shell NCs with disordered metallic structures, exhibiting very localized plasmon modes. In contrast with many previous experiments \cite{Ito07,Matsuda08,Schietinger09,Wu10,Vion10}, there is no excitation enhancement near the metal film which could contribute to the fluorescence enhancement. The coupling with the metal plasmons is solely due to the QD emission, resulting in $F_P$ values as high as 60. The analysis of the photon's time statistics demonstrates that strong antibunching or biexcitonic cascades can be achieved. The Auger processes, modified by the CdS's shell thickness, and the acceleration of the radiative processes bring about the strong antibunching or biexcitonic cascades. Complete suppression of the blinking is also reported. Finally, it will be shown that a very high efficiency of the photon collection is obtained. Up to 37 \% (a value twice as high as that for a glass coverslip) of the propagative photons emitted during the decay processes are collected through the microscope objective.

The CdSe/CdS NCs ($\lambda$=660 nm) are composed of a CdSe core (3 nm radius) and a giant crystalline shell of CdS (10 nm, rms $\sim$ 1 nm). Their synthesis is described in ref. \cite{Mahler08}. In quasi-type II CdSe/CdS NCs, the electron is partially delocalized in the shell. Hence Coulomb interactions are reduced resulting in an increase of the lifetimes of Auger processes. Since these processes control the quantum properties of the emission, this increase has important consequences. In contrast with standard CdSe/ZnS NCs, the fluorescence of charged CdSe/CdS NCs is not quenched by the Auger processes. Under low power pulse excitation, the probability to generate one electron-hole (e-h) pair per pulse is much higher than that to create several. In this case, CdSe/CdS NCs oscillate between two states \cite{Spinicelli09}, a bright and a grey one. The bright state corresponds to the monoexciton radiatively recombined (neutral NC), and the grey state corresponds to the recombination of the trion (ionized NC) that occurs with a lower radiative QE (see Supplemental Material). 

The random gold film is prepared just below the percolation threshold \cite{Ducourtieux01} (see supplemental material). It is characterized by a wide distribution of plasmon resonances. They cover the visible spectrum above 550 nm (see Fig. \ref{fig2}.b). Thus the emission of the NCs is within the plasmon resonances \cite{Buil06}. In these disordered films, high localization of plasmon modes has been identified. The electromagnetic energy is confined in subwavelength-sized regions called ``hot spots''. A strong enhancement of the intensity has been theoretically predicted and experimentally observed \cite{Ducourtieux01,Buil06}. These localized surface plasmons have also recently been identified due to the fluctuations of the local density of optical states \cite{Krachmalnicoff10}.

The fluorescence of individual NCs is analyzed with a confocal microscope and a Hanbury-Brown and Twiss setup (PDM photodiodes, time resolution = 50 ps, Picoquant PicoHarp300). 
A pulsed laser diode (Picoquant, $\lambda$ = 485 nm, pulses duration = 100 ps) provides the optical excitation. This setup captures the absolute time of arrival of photons with a 64 ps accuracy. The time evolution of the intensity, the histogram of the delays between photons pairs as well as the photoluminescence (PL) decay can be analyzed. An arbitrary range of intensity of the photons can also be selected in order to extract the abovementioned properties for the neutral and ionized states of the NC \cite{Spinicelli09}.

\begin{figure}[t]
\includegraphics[width=15cm]{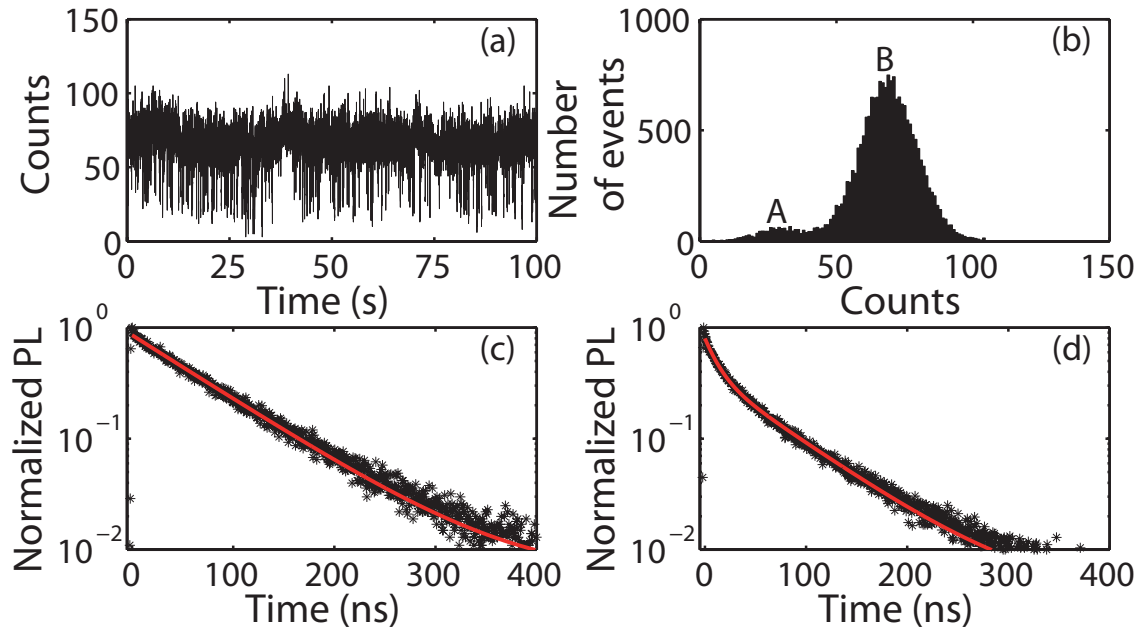}
\caption{(a) Fluorescence intensity of a single CdSe/CdS NC on a glass coverslip ($t_b =$ 10 ms). The background level is 0.3 count per time bin and can be neglected. (b) Histogram of the intensity corresponding to (a). (c) PL decay of the bright state (peak B). The red line is a monexponential fit (with an offset, lifetime of 75 ns). (d) PL decay of the low emitting periods (bump A). The red line is a bi-exponential fit (lifetimes of 69 ns and 12 ns).} \label{fig1}
\end{figure}

\begin{figure}[t]
\includegraphics[width=15cm]{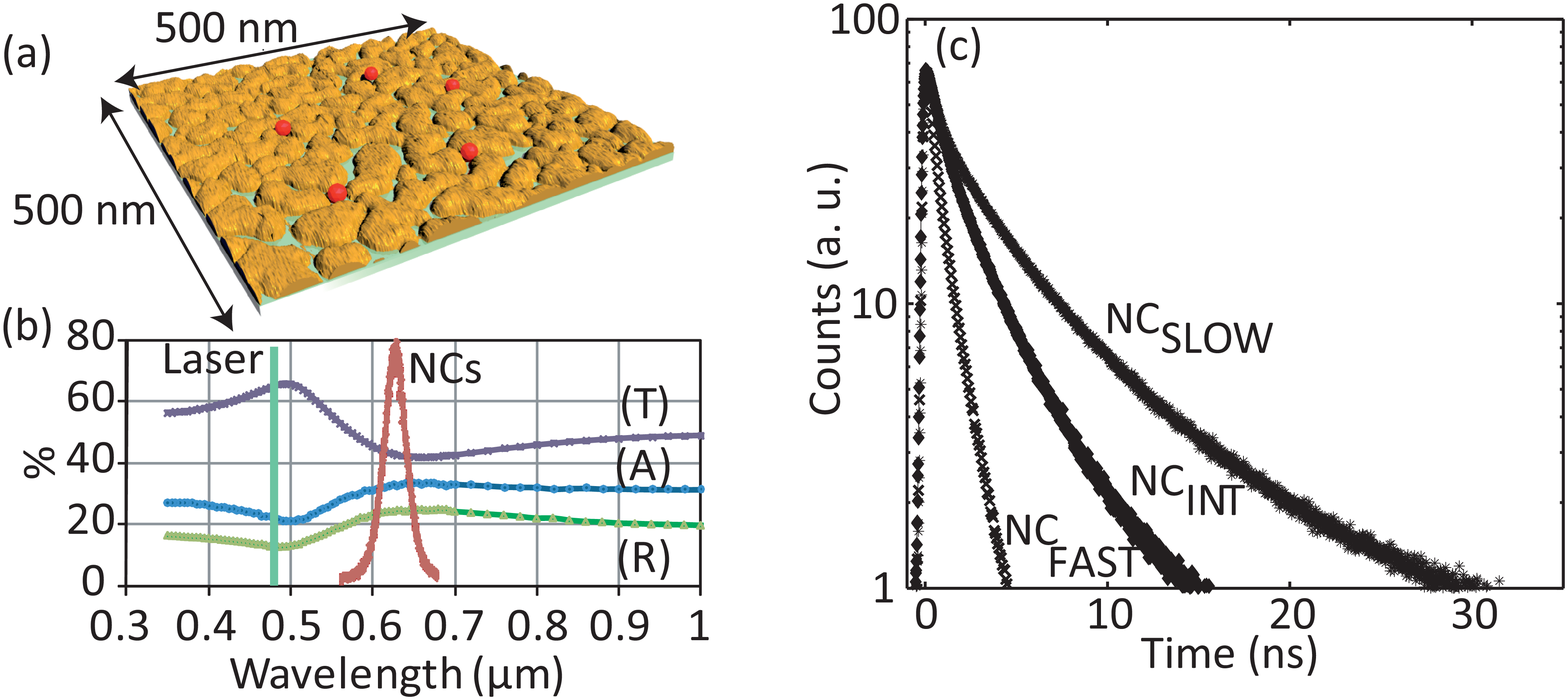}
\caption{(a) Sketch of NCs deposited on the semicontinuous gold film. The topography of the gold film has been realized with an AFM. NCs are represented to scale. (b) Absorbance, (A), transmittance (T) and reflectance (R) of the random gold film. The laser diode wavelength (Laser) is represented as well as the NCs photoluminescence (NCs). (c) PL decay of three NCs illustrating the range of decay rates. The PL decay of NC$_{\rm FAST}$ is very fast. NC$_{\rm INT}$ corresponds to an intermediate case. For the NC labeled NC$_{\rm SLOW}$, the decay is relatively slow.} \label{fig2}
\end{figure}

To begin, the emission of individual NCs on a glass coverslip (oil objective, N.A. = 1.4) is characterized. As in ref. \cite{Mahler08,Spinicelli09}, a perfect antibunching (see Supplemental Material) was always observed confirming that the studied NCs are individual. The radiative lifetime of the bright state is found to be around $\tau_{ref}$= 75 ns (Fig. \ref{fig1}c). The histogram of the intensity shows that the occurence of the grey state is very low (Fig. \ref{fig1}b). A QE of 40 \% for the grey state is deduced from the relative positions of the A and B bumps in Fig. \ref{fig1}b (see Supplemental Material). The PL decay during grey periods is well adjusted by a bi-exponential. The short component provides the trion lifetime (12 ns). The long component of 69 ns is very close to $\tau_{ref}$.
Indeed, due to the finite duration of the time bin ($t_b)$, photons emitted during bright states contribute to the intensity of low emitting periods. From the method described in \cite{Spinicelli09} (see Supplemental Material), the Auger lifetime ($\sim$ 20 ns) and the radiative lifetime ($\sim$ 30 ns) of the trion state are deduced. These values are used as references in the following.

When an emitter is coupled to a metallic structure, radiative and non radiative channels are strongly modified. Their relative importance depends crucially on the distance between the fluorophore and the metal. Intensity enhancement as well as quenching is observed \cite{Matsuda08}. The distance between the metal and the emitter can be controlled through various approaches such as the use of a silica spacer \cite{Mallek10} or biomolecules \cite{Gueroui04}. A previous study has also shown results concerning NCs being deposited on the metal itself \cite{Ito07}. However, in our paper, the crucial role of a spacer played by the thick shell of the NC is demonstrated. In constrast with our previous study \cite{Mallek10} concerning CdSe/CdS NCs with a thiner shell (5nm), no silica spacer is needed to avoid quenching of the fluorescence.

The relative part of radiative versus non radiative processes can be studied through the collection efficiency of far field photons. Several references are needed to determine these fundemental parameters. The emission of individual NCs on a glass coverslip is collected through an air objective (N.A. = 0.95, always used in the following experiments). At saturation, the bright states correspond to a single photon emission. The collection efficiency is then equal to 0.4 \% i.e. 0.4 \% of the laser pulses generate a photon that is detected by the optical setup. Considering the transmission (2 \%) of the optical setup excluding the objective, a collection efficiency of about 20 \% of the air objective is calculated. This value is consistent with theoretical predictions \cite{Brokmann05}. In the following, the pump power is set to the value $\mu$ corresponding to a fluorescence intensity ten times lower than the intensity at saturation. The probability to excite several e-h pairs by one pulse is very low. The probability $P(n)$ to generate $n$ e-h pairs per pulse follows a Poissonian statistics \cite{Spinicelli09}, and thus $P(n) = \eta^n \exp(-\eta)/ n!$ ($\eta$ is the average of the distribution). $\eta$ = 10.5 \%, $P(1)$ = 9.5 \% and P(2) = 0.5 \% are deduced from the  value $\sum_{1}^{\infty} P(n)= 10\%$. The probability to excite several e-h pairs is about 20 times lower than the probability to excite one. This probability is not modified by the random gold film. Indeed, the reflectivity of the film at the laser wavelength is very low (13 \%, see Fig. \ref{fig2}.b). No plasmon resonances are excited and the fluctuations of the near-field intensity at 485 nm are lower than 10 \% \cite{Buil06}. The excitation power does not depend on the NC position and is assumed to be equal to that of the glass coverslip. This property is a key point since it enables us to control the pump power and to make a precise assessment of the collected photons under pulsed excitation. Moreover, the fluorescence enhancement described below is solely due to the coupling of the NC emission with the gold film and is not obtained, even partially, through excitation enhancement as described in many works \cite{Ito07,Matsuda08,Schietinger09,Wu10,Vion10}.

The fluorescence of NCs directly deposited on the metal film will be now investigated. A drastic modification in the decrease of the NCs PL is reported. For more than 50 NCs, the PL decay at 1/e ranges from 0.8 to 2.9 ns. When compared to $\tau_{ref}$, these very low values show a strong Purcell effect. Fig. \ref{fig2}.b represents the results corresponding to 3 NCs. The NC labeled NC$_{\rm FAST}$ is a typical NC with a very fast decay ($\sim$ 0.8 ns) while the NC labeled NC$_{\rm SLOW}$ is representative of NCs with a relatively slow decay ($\sim$ 2.9 ns). NC$_{\rm INT}$ represents an intermediate case. The large dispersion in the decay rates that is observed can be attributed to the random spatial localization of the plasmons on the disordered gold film \cite{Buil06}. This dispersion has already been observed for NCs deposited on such films covered with a 30 nm silica spacer \cite{Mallek10}.

The histogram of the delays between photons is also strongly modified. For NC$_{\rm FAST}$ (Fig. 3a), the peak at zero delay has the same areas as the other ones (large fluctuations are observed in the peak heights but not in the peak areas). This NC is no more a single photon emitter. The biexcitonic state recombines radiatively. Bearing in mind that, due to low pumping, multiexcitonic states of higher order can be neglected. The radiative recombinations are drastically accelerated being faster than the Auger ones. 
\begin{figure}[t]
\includegraphics[width=15cm]{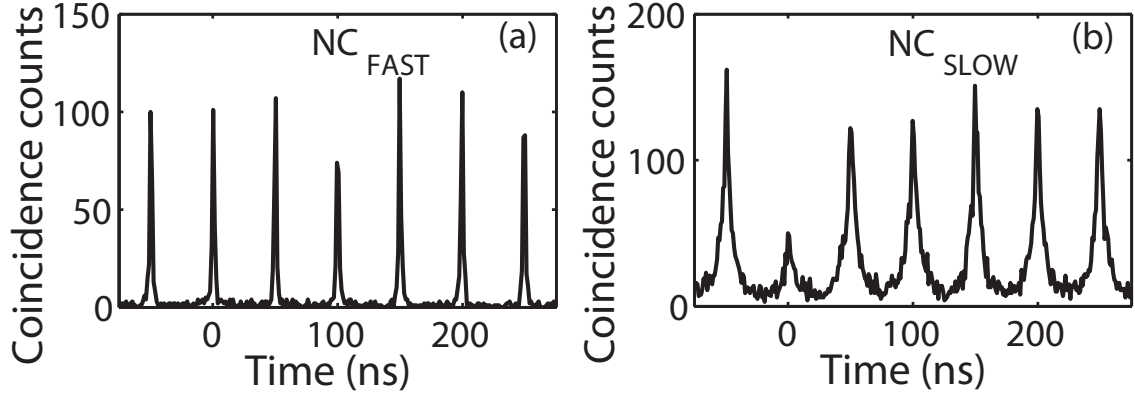}
\caption{Coincidence counts for NC$_{\rm FAST}$ (a) and NC$_{\rm SLOW}$ (b).} \label{fig3}
\end{figure}

As stated before, the pump power $\mu$ corresponds to a low probability to create several e-h pairs. However, the peak at zero delay has the same area as the other ones. This can be deduced from the Poissonian statistics followed by $P(n)$. Since $\eta = 0.105 \ll 1$, $P(n)$ can be approximated by $P(n) = \eta ^n/n!$, leading to $P(1)=\eta$ and $P(2) = \eta^2/2$. Radiative recombination of all pairs is assumed. The peak at zero delay corresponds to the probability to detect two photons emitted by the NC after a single laser pulse. This area is proportional to $P(2)/2 = \eta ^2 /4$ (the divider 2 accounts for the Hanbury-Brown and Twiss setup). At low excitation, the peak at the delay $T$ ($T \neq 0$) corresponds to the probability to detect two photons generated by two pulses separated by $T$. This area is proportional to $P(1)/2 \times P(1)/2 = \eta ^2 /4$. Despite the weak probability to create two e-h pairs ($P(2)/P(1) = 20$), the peak at zero delay has the same area as the other ones. For NC$_{\rm FAST}$, the probability that the biexciton recombination leads to the collection of two photons is the same as the probability that the monoexciton recombination induces the collection of one photon. In addition to the Auger processes suppression, Fig. \ref{fig3}a shows that the biexcitonic and monoexcitonic states are equally coupled to the gold film. This is due to the spectral width of a plasmon mode ($>$ 30 nm \cite{Ducourtieux01}) that is much larger than the total linewidth of the NC emission (15 nm). In the case of NC$_{\rm FAST}$, biexcitonic cascades are observed. 

In strong contrast, the peak at zero delay for NC$_{\rm SLOW}$ is smaller than the other ones (Fig. \ref{fig3}b). Auger processes dominate radiative ones. Strong antibunching (64 \%) is observed.

\begin{figure}[t]
\includegraphics[width=15cm]{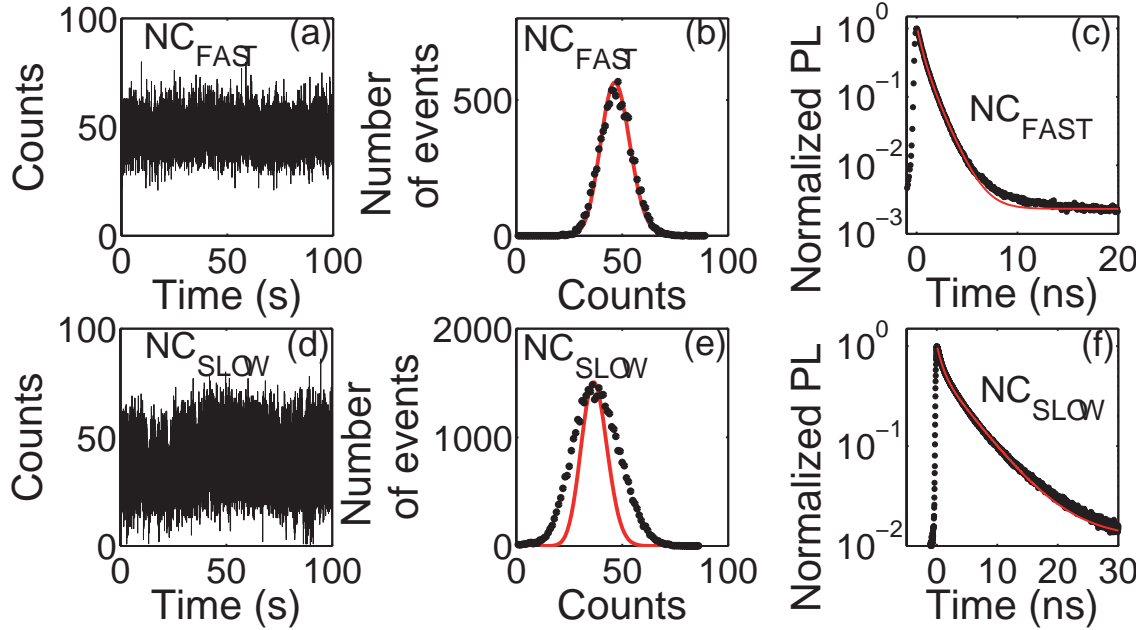}
\caption{(a) Fluorescence intensity of NC$_{\rm FAST}$ ($t_b =$ 10 ms). (b) Histogram of the intensity corresponding to (a). The dots are the experimental results. The red line is a Poissonian distribution fit (mean value of 47). (c) PL decay of NC$_{\rm FAST}$. The red line is a bi-exponential fit (lifetimes of 0.49 ns and 1.28 ns). (d) Fluorescence intensity of NC$_{\rm SLOW}$  ($t_b =$ 2.5 ms). (e) Histogram of the intensity corresponding to (d). The dots are the experimental results. The red line is a Poissonian distribution fit (mean value of 37). (f) PL decay of NC$_{\rm SLOW}$. The red line is a bi-exponential fit (lifetimes of 0.9 ns and 5.49 ns).} \label{fig4}
\end{figure}

Fig. \ref{fig4} presents the fluorescence intensity of NC$_{\rm FAST}$ and NC$_{\rm SLOW}$ and the histogram of the intensity. For a NC on a glass coverslip, the histogram shows two states corresponding to the grey and bright ones (Fig. \ref{fig1}b). On the random gold film, the shape of the curve is modified. Moreover, for NC$_{\rm FAST}$, no blinking nor grey states are observed, in contrast with the results presented in \cite{Mallek10} where a silica spacer is used. More quantitatively, the histogram of the intensity is very well adjusted by a Poissonian distribution. The NC$_{\rm FAST}$ no longer oscillates between the two states, each of different QE. Two explanations are possible. Firstly, the radiative recombinations have become faster than the Auger recombination and the QE of the trion state is equal to that of the neutral state. Secondly, NC$_{\rm FAST}$ has exchanged an electron with the metallic surface and remains ionized. The emission is always due to the trion recombination and is associated to only one state. 

The histogram of the intensity for NC$_{\rm SLOW}$ (Fig. \ref{fig4}e), is no more well fitted by a Poissonian distribution. The QE of ionized and neutral states are similar but not equal. Radiative processes are slower and the Auger recombination is sometimes faster than the radiative ones. The trion has a lower QE than the neutral exciton. The fit of the PL decays for NC$_{\rm FAST}$ and NC$_{\rm SLOW}$ (Fig. \ref{fig4}c and Fig. \ref{fig4}f) reinforces the interpretation assuming two states. For a minimum of two decades, the PL decay of the two NCs is very well fitted by a bi-exponential curve. These decay rates can be interpreted as the rates of the neutral and ionized states.
In addition, the blinking of the QDs emission is suppressed, but for reasons fundamentally different than the one evocated in \cite{Ito07}. Here, we are using thick shell QDs that have strongly suppressed Auger recombination \cite{Spinicelli09}. Radiative lifetimes shorter than the auger recombination implies that QDs will be bright even if charged. This result is completely different from the assumption that the blinking is suppressed because the QDs are still in neutral state \cite{Ito07}.

The $F_P$ factor for the neutral state can also be calculated from these decay rates. $F_P$ is defined as $\tau_{ref}/\tau_b $ ($1/\tau_b$ is the total decay rate of the bright state, that includes the radiative and non radiative contributions) \cite{Esteban10}. For NC$_{\rm SLOW}$, a $F_P$ factor of 14 is calculated while, for NC$_{\rm FAST}$, $F_P$ can be as high as 60. These values demonstrate the dramatic increase of the decay rate of the monoexcitonic state. This increase could mainly correspond to the opening of non radiative channels. The energy would be transferred to the gold film and dissipated, resulting in a strong quenching of the NCs' fluorescence. On the contrary, the following results concerning the collection efficiency will show that plasmons are effectively coupled to the far field by the film structure.

The relative part of the radiative versus non radiative processes leading to the desexcitation of the monoexcitonic state are determined by studying the photon collection. Values of 4.7 kHz for NC$_{\rm FAST}$ and 14.8 kHz for NC$_{\rm SLOW}$ are deduced from the intensity histograms (pulse repetition rate = 20 MHz). For a NC on a glass coverslip, the intensity is 2 kHz (laser repetition rate = 5 MHz) and corresponds to a collection efficiency of the objective of 20 \%. Since the probability to excite the NC per pulse is equal in both configurations, a precise assessment of photons collected can be made. The total decay rate $k_n$ is the sum of $k_{rad,coll}$ and $k_{other}$ ($k_{rad,coll}$ is the decay rate corresponding to photons emitted in the far field and collected by the objective, $k_{other}$ is the decay rate corresponding to other processes - non-collected photons, non-radiative processes, non-scattered plasmons). The probability to excite the NC (10 \%), the repetition rate (20 MHz) and the total optical transmission (2 \%) excluding the objective, give the percentage of collected photons. Due to low power excitation, the probability to generate two photons by one pulse is negligible (even in the case of NC$_{\rm FAST}$). For NC$_{\rm SLOW}$, the percentage of photons collected is equal to 37 \% ($14.8 \times 10^3/(20 \times 10^6 \times 0.02 \times 0.1)$). For NC$_{\rm FAST}$, the result is 12 \%. These values are of the order or higher than in the case of a glass coverslip.

Each decay rate for a NC deposited on a glass coverslip or on a random gold film (Tab. \ref{tab1}) is deduced from the total decay rate $k_n$ and from the fraction of photons collected $k_{rad,coll}/k_n$. The results of Tab. \ref{tab1} show that a large number of radiative channels are opened by the coupling of NCs on gold film. This number largely exceeds the number of radiative channels for a NC on a glass coverslip. As for the total decay rate, the number of opened radiative channels depends on the position of the NC. The percentage of photons collected appears to be correlated to the decay rate: the increase of coupling of NCs on gold film induces a decrease in the percentage of photons collected. However, the emission of NCs can never be quenched, thus it can be assumed that the giant shell acts as a spacer. The shell (thickness = 10 nm) is thick enough to avoid quenching as theoretically predicted \cite{Lakowicz05,Mertens07}. A non quenching quality is another crucial characteristic of these NCs.
\begin{table}[h]
\begin{center}
\begin{tabular}{ c c c c c c }
 & Lifetime  & I & \% of collected & k$_{rad,coll}$ & k$_{other}$ \\
 & (ns) & (kHz) & photons & (MHz) & (MHz) \\
\hline
NC$_{\rm SLOW}$  & 5.49 & 14.8 & 37 & 68 & 115 \\
NC$_{\rm INT}$ & 3.47 & 8.8 & 22 & 65 & 231 \\
NC$_{\rm FAST}$ & 1.28 & 4.7 & 12 & 92 & 689 \\
Reference & 75 & 2 & 20 & 2.7 & 10.7 \\
\end{tabular}
\end{center}
\caption{PL lifetime and decay rates of the 3 NCs corresponding to Fig. \ref{fig2}. The line ``Reference'' corresponds to the mean results obtained for NCs deposited on a glass coverslip.} \label{tab1}
\end{table}

In conclusion, the fluorescence of giant shell CdSe/CdS NCs directly deposited on a random gold film has been studied in detail. The ``hot spots'', generated by the disorder, induce strong Purcell effects which results in an almost complete suppression of blinking. The quantum properties of the emission are determined by the strength of the NC/random film coupling. Strong antibunching or biexcitonic radiative cascades are observed with a very large collection efficiency. In the field of quantum plasmonics, this paper highlights the great interest of coupling thick shell NCs with metallic structures. 

Acknowledgement: B. D. and J.-P. H. thank R\'egion Ile de France and Agence Nationale de la Recherche fundings. All the authors thank F. Vega for fruitful discussions and advices.

\end{document}


\section{Fluorescence properties of the CdSe/CdS NCs deposited on a glass coverslip}
\begin{figure}[ht]
\includegraphics[width=15cm]{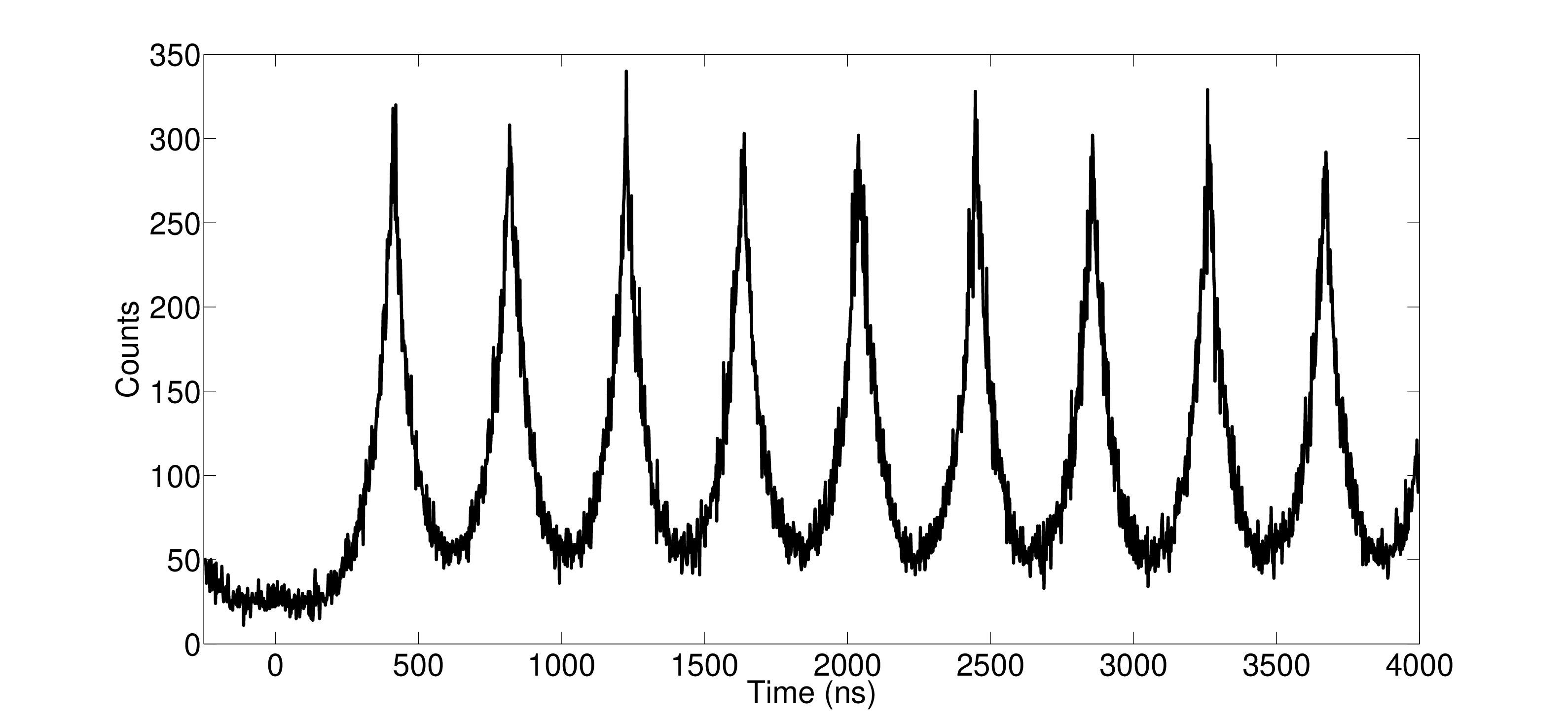}
\caption{Coincidence counts for a typical NC deposited on a glass coverslip.} \label{fig1_SM}
\end{figure}

The very thick shell NCs used in the experiments were synthesized and characterized in detail by the authors. Their structural and optical properties are very well controlled. At the pump rate used in the experiments, single photon emission is always observed when NCs are deposited on a glass coverslip (see Fig. \ref{fig1_SM}). 

\begin{figure}[ht]
\includegraphics[width=10cm]{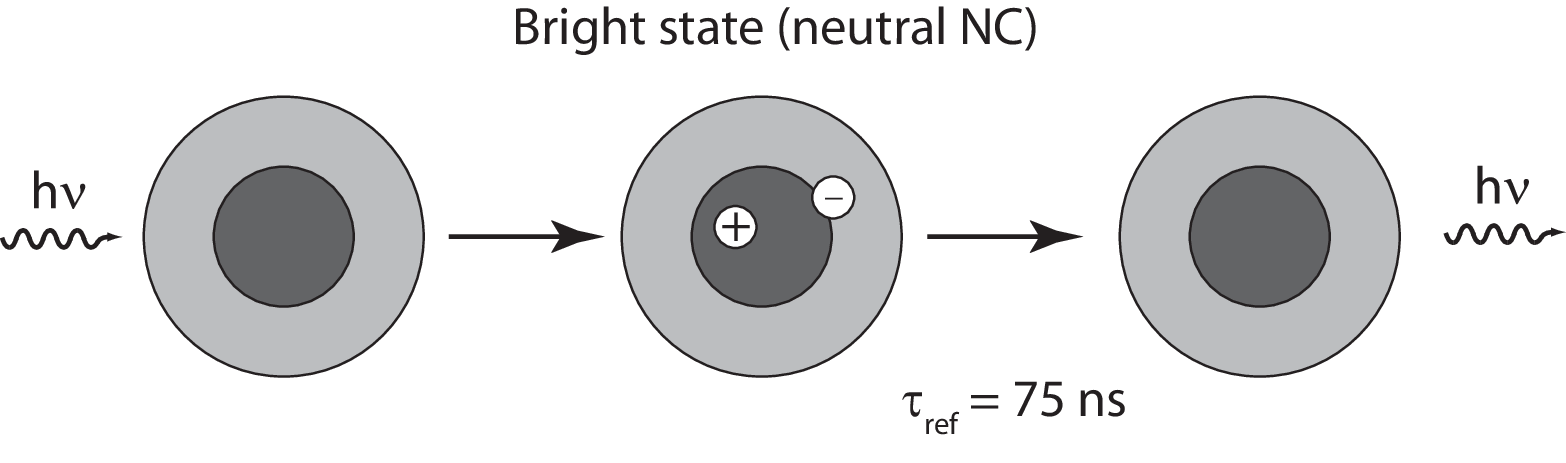}
\caption{Schematic representation of the bright state} \label{fig2_SM}
\end{figure}

Their fluorescence alternate between two states. The bright state corresponds to a neutral NC (Fig. \ref{fig2_SM}). In this case, the monoexcitonic state always recombines radiatively . As for standard CdSe/ZnS NCs \cite{Brokmann04b}, the QE efficiency of the bright state of CdSe/CdS NCs has been measured very close to unity \cite{Spinicelli09}. Its radiative lifetime $\tau_{ref}$ is around 75 ns (Fig. 1c of the article). 

\begin{figure}[ht]
\includegraphics[width=10cm]{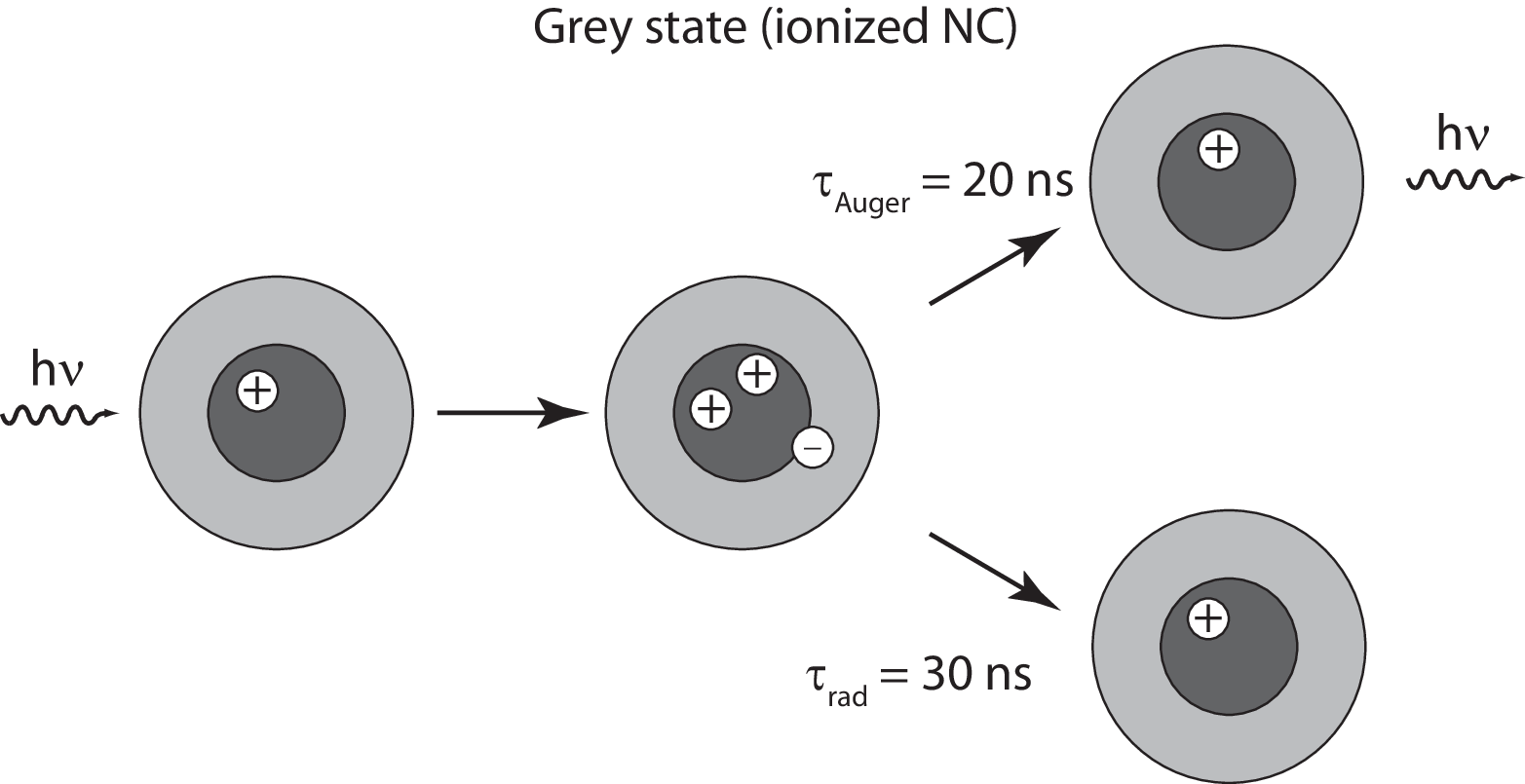}
\caption{Schematic representation of the grey state} \label{fig3_SM}
\end{figure}

The grey state was attributed to the ionized state of the NC. In quasi-type II CdSe/CdS NCs, the hole is confined in the core. In contrast, the electron is delocalized in the whole structure due to relative positions of CdSe and CdS conduction bands. For the trion state, two recombination processes have to be considered: an electron hole-pair can recombine radiatively (rate $k_{rad}$) or through an Auger process (rate $k_{A}$), the energy being then transfered to the remaining charge (Fig. \ref{fig3_SM}). Indeed ionization of the NC does not modify its close environment and does not create new non radiative pathways except the Auger one. The total desexcitation rate $k$ is the sum $k_{rad}+k_{A}$ and corresponds to the short component of the total PL decay (12 ns).

Following the method of \cite{Spinicelli09}, from the histogram of the intensity (Fig 1.b of the manuscript), the QE of the trion state $Q=k_{rad}/(k_{rad}+k_{A})$ can be deduced. It is equal to the ratio between the intensities of the grey and bright states  which are given by the relative positions of bumps A (28 counts / 10 ms) and B (68 counts / 10 ms) in Fig 1.b. Since the QE of the bright state is close to 100 \%, the QE of the grey state is about 40 \%. 

From $Q=k_{rad}/(k_{rad}+k_{A})=40$ \% and $k=k_{rad}+k_{A}= 1/12$ ns$^{-1}$, a straightforward calculation gives $k_{rad}=1/30$ ns$^{-1}$ and $k_A=1/20$ ns$^{-1}$.

\section{Properties of the random gold film}
Au films characteristics depend strongly on the experimental conditions and the substrate. However, the metallic structure used in the experiments is a random gold film at the percolation threshold deposited on a glass coverslip, It is realized by the authors and studied in detail by several groups including ours (\cite{Ducourtieux01,Stockman01,Krachmalnicoff10,Buil06}). It is prepared by evaporation under vacuum (10$^{-9}$ torr). The mass thickness is 61 $\AA$. A typical AFM image is shown in Fig. 2.a of the paper. Its optical properties are very well known. In particular, the disorder is at the origin of the strong enhancement of the electromagnetic field in subwavelength-sized regions. The corresponding plasmon resonances cover the visible range above 550 nm as shown by the absorption spectra of Fig. 2.b. This figure also illustrates the originality of our approach: the excitation ($\lambda$ = 485 nm) cannot be enhanced by the excitation of plasmons of the random gold film. Only the NCs emission is coupled to plasmon resonances of the gold structure. The properties of the metallic structure as well as those of the nanocrystals have been mastered in this work.